# On the Hardness of a New Boron Phase, Orthorhombic γ-$B_{28}$

V. L. Solozhenko[a], O. O. Kurakevych[a], and A. R. Oganov[b]

[a]*LPMTM—CNRS, Université Paris Nord, France*
[b]*Department of Materials, ETH Zürich, Switzerland*

**Abstract**—Measurements of the hardness of a new high-pressure boron phase, orthorhombic γ-$B_{28}$, are reported. According to the data obtained, γ-$B_{28}$ has the highest hardness (~ 50 GPa) of all known crystalline modifications of boron.

Among the 16 polymorphous modifications of boron mentioned in the literature [1] only three phases seem to correspond to the pure element. They are: the α-$B_{12}$ rhombohedral low-temperature [2], β-$B_{106}$ rhombohedral high-temperature low-pressure [3], and t-$B_{192}$ tetragonal high-temperature [4] phases. Recently a new high-pressure boron phase, namely, orthorhombic γ-$B_{28}$ [5] has been synthesized. The structure of this phase (Fig. 1a) has been established [5] by *ab initio* calculations in the framework of the USPEX evolutionary algorithm [6]. At present only the hardness of the α-$B_{12}$ ($H_V$ = 42 GPa [7]) and β-$B_{106}$ ($H_V$ = 45 GPa [8]) phases has been measured and the experimental values are in good agreement with the values (39.2 and 43.8 GPa, respectively) calculated by us in the framework of the thermodynamic models of hardness [9, 10]. In the present study the hardness of polycrystalline orthorhombic γ-$B_{28}$ is measured for the first time.

As the initial material for synthesis of γ-$B_{28}$, highly crystalline β-$B_{106}$ (99.995 at %) was used. The γ-$B_{28}$ phase was synthesized in a multianvil two-stage apparatus (a 1000-ton press Max Voggenreiter/Walker module) at pressures from 15 to 20 GPa and temperatures from 1800 to 2000 K for 10–30 min. To isolate the sample from the high-pressure cell elements, capsules of pyrolytic boron nitride, which does not react with crystalline boron at temperatures below 2000 K [11], have been used. According to X-ray diffractometry (TEXT 3000, INEL) and elemental analysis (SX-50 Camebax, Cameca), the prepared samples are defined as crystalline γ-$B_{28}$ (the orthrhombic system, space group *Pnnm*, $a$ = 5.054 Å, $b$ = 5.612 Å, $c$ = 6.966 Å) that does not contain impurities.

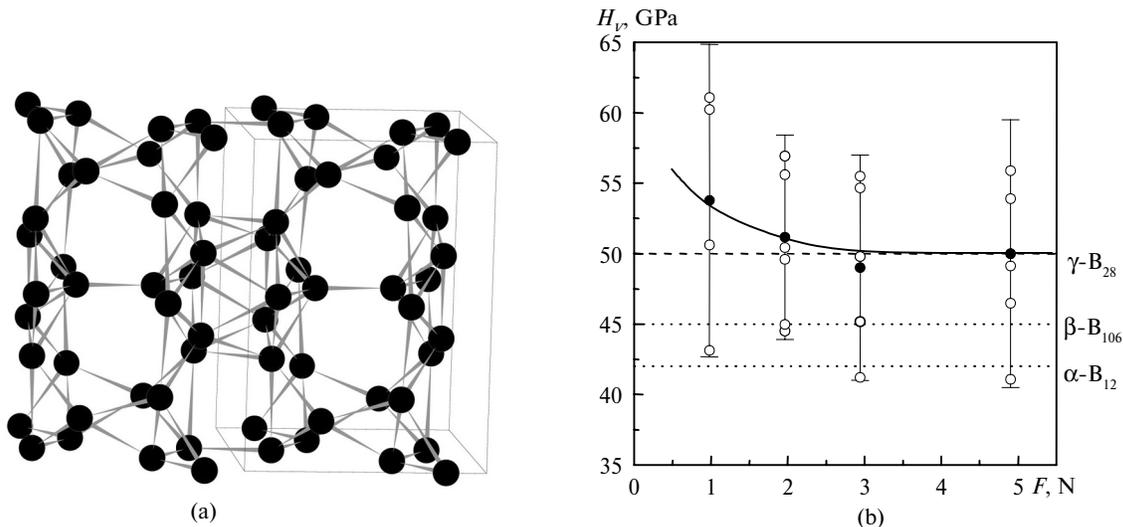

**Fig. 1.** The γ-$B_{28}$ phase crystal structure (a) and the phase hardness vs. load (b). The centres of gravity of $B_{12}$ icosahedra and $B_2$ dumbbells in the structure of the new phase form a NaCl-type structure [5].

The Vickers hardness was measured using a Struers Duramin-20 microhardness tester under loads ($F$) from 1 to 20 N and at an indentation time of 20 s. At least 4 indents were spaced at about 200 µm intervals. At $F > 10$ N the appearance of long cracks and spallings obstructed accurate estimation of the hardness.

According to the data obtained (Fig. 1b), polycrystalline phase has a hardness of 50 (11) GPa, which is markedly higher than the hardness of other boron modifications [7, 8] and agrees well with the value of 48.8 GPa that was calculated in the framework of the thermodynamic model of hardness [9, 10]. Thus, new high-pressure phase $\gamma$-$B_{28}$ has the highest hardness among the known boron crystalline modifications, which stems from its highest density (2.544 g/cm$^3$).

The authors are grateful to Jiuhua Chen and Yanzhang Ma, who independently synthesized the $\gamma$-$B_{28}$ phase and participated in the determination of the phase structure. We also thank the Agence Nationale de la Recherche (grant ANR-05-BLAN-0141) and Swiss National Science Foundation (grants 200021-111847/1 and 200021-116219) for financial support.